\documentstyle[multicol,aps,prl]{revtex}

\begin{document}

\widetext

\draft

\title{Chaotic flow and efficient mixing in microchannel with a polymer solution.}

\author{Teodor Burghelea$^a$, Enrico Segre$^b$, Israel Bar-Joseph$^c$, Alex Groisman$^d$ and
Victor Steinberg$^a$}
\address{$^a$ Department of Physics of Complex Systems, Weizmann Institute of Science, Rehovot, 76100 Israel;\\
        $^b$ Department of Physical Services, Weizmann Institute of Science, Rehovot, 76100 Israel;\\
         $^c$ Department of Condensed Matter Physics, Weizmann Institute of Science, Rehovot, 76100 Israel;\\
          $^d$ Department of Physics, UCSD, 9500 Gilman Dr., La Jolla, CA, 92093-0374.}

\date{\today}

\maketitle
\begin{abstract}

Microscopic flows are almost universally linear, laminar and stationary because Reynolds number, $Re$,
is usually very small. That impedes mixing in micro-fluidic devices, which sometimes limits their
performance. Here we show that truly chaotic flow can be generated in a smooth micro-channel of a
uniform width at arbitrarily low $Re$, if a small amount of flexible polymers is added to the working
liquid. The chaotic flow regime is characterized by randomly fluctuating three-dimensional velocity
field and significant growth of the flow resistance. Although the size of the polymer molecules
extended in the flow may become comparable with the micro-channel width, the flow behavior is fully
compatible with that in a table-top channel in the regime of elastic turbulence. The chaotic flow
leads to quite efficient mixing, which is almost diffusion independent. For macromolecules, mixing
time in this microscopic flow can be three to four orders of magnitude shorter than due to molecular
diffusion.
\end{abstract}
\pacs {PACS numbers: 47.27.-i,47.50.+d,83.50.-v}

\begin{multicols}{2}
\narrowtext

\section{Introduction.}

Flows of liquids in microscopic channels have been attracting increasing interest recently due to fast
development of microfluidics and soft lithography\cite{1,2}. The microfluidic devices, which are
becoming increasingly advanced and reliable, allow dramatic reduction of amounts of reagents required
for fine chemistry and biochemistry\cite{3}, well controlled manipulation and sophisticated
experiments on individual cells\cite{4,5,6} and macromolecules\cite{7}. The microscopic flows are
almost universally laminar, with linear dependence of the flow rate on the driving force. They also
usually remain steady as long as the driving force does not change. All of that has to do with low to
moderate values of the Reynolds number, $Re=Vd\rho/\eta$, which is a general measure of non-linear
inertial effects in the flow and of likelihood to find it being chaotic or turbulent\cite{8}. Here $V$
is the flow velocity, $d$ is the diameter of the channel, and $\rho$ and $\eta$ are the density and
the viscosity of the fluid, respectively. When $d$ is reduced, the flow velocity needed to reach a
given high $Re$, which is required to generate chaotic or turbulent flow, increases as $d^{-1}$ . The
driving pressure per unit length scales like $\Delta P/\Delta L\sim \eta V/d^2$ giving $\Delta
P\sim\Delta L/d^3$ at a given $Re$ in the channel. Therefore, if the channel proportions are
preserved, $\Delta P$ grows quadratically with $d^{-1}$ and when the channels are only a few tens of
microns wide, achieving sufficiently high $Re$ requires impractically high driving pressures.

The laminar character of microscopic flows has multiple practical advantages including possibility of
precise control of flow velocity, chemical concentration profiles\cite{4,5,6}, and targeted delivery
of chemicals and particles\cite{5}. On the other hand, the laminar flows have an inherent problem of
inefficient mass transfer in directions perpendicular to the main flow, which occurs due to molecular
diffusion only. Diffusion time, $d^2/D$, across a typical micro-channel with a width of 100 $\mu$m is
on the order of 100 s even for moderate size proteins, such as bovine serum albumin with a diffusion
coefficient of $D\approx 3\cdot 10^{-7}$ cm$^2$/s in water\cite{9}.

A few techniques have been proposed to generate stirring by a three-dimensional flow in order to
increase the rate of mixing in the micro-channels. They include application of time dependent external
forces\cite{10,11} and raising $Re$ to moderately high values in curvilinear three-dimensional
channels\cite{12,13}. An ingenious method of mixing has been suggested recently, which involves
special "herring-bone" patterning of a micro-channel wall to generate fluid motion perpendicular to
the main flow direction in the linear, low $Re$ regime\cite{14}. The fluid elements are continuously
stretched and folded in the flow as they advance along the channel. That separates closely spaced
fluid particles and brings distant particles together, dramatically reducing the characteristic length
scales and diffusion time and increasing the rate of homogenization of the mixture. The flow was
stationary in the laboratory frame, however. Therefore, the concentration profiles of the  fluorescent
dye used as a tracer were uniquely defined by the entrance conditions and the channel
geometry\cite{14}. They did not change in time, and there was always some constant difference in
concentration between neighboring points.

The basic condition of linearity in low $Re$ flows can be changed by adding flexible high molecular
weight polymers to the working liquid\cite{15}. Solutions of those polymers are known as non-Newtonian
visco-elastic fluids\cite{15}. Mechanical stress in these fluids depends on the flow history with some
characteristic relaxation time, $\lambda$, which for dilute solutions is a time of relaxation of
individual polymer molecules. Another specific property of the polymer solutions is the non-linear
dependence of the polymer contribution to the stress on the rate of deformation in the flow, $\nabla
V$ \cite{15}. This non-linearity usually becomes significant, when the Weissenberg number,
$Wi=\lambda\nabla V$, becomes on the order of unity. The non-linear growth of the elastic polymer
stresses is especially striking in extensional flows at $Wi>1/2$, where apparent viscosity of polymer
solutions can rise by up to three orders of magnitude as the total deformation increases\cite{16}. In
pure shear flows the major non-linear elastic effect shows up in the appearance of negative normal
stress along the flow direction. It leads to the well known effect of rod climbing\cite{15,17} and
causes purely elastic instabilities in curvilinear flows of viscous polymer solutions in table-top
set-ups, when inertial effects are virtually absent\cite{18}. The non-linear polymer stresses in
curvilinear shear flows can also lead to elastic turbulence, a random multi-scale three-dimensional
flow, which can develop at arbitrarily low $Re$\cite{19}. Elastic turbulence causes sharp growth of
the flow resistance\cite{19,20}, and it was found to generate efficient mixing in a table-top
curvilinear channel\cite{21}.

Recently it was shown that the non-linear elasticity of the polymer solutions can cause a purely
elastic transition and non-linear growth of flow resistance in a microscopic channel with
contractions\cite{22}. Those non-linear effects, however, were due to regions of fast extensional flow
near the contractions, where individual polymer molecules are expected to partially unravel at
$Wi>1/2$ \cite{23,24}. The non-linearity in resistance was the most dramatic feature of the transition
and it became quite significant before any substantial changes in the flow pattern could be
seen\cite{22}. Therefore the non-linear resistance could be understood as a simple additive effect of
individual molecules forced through the contractions. Although the flow became rather irregular at
higher flow rates, the fluctuating flow regions were mostly near the contractions, and no detailed
study of those fluctuations was made.

If the basic linear flow is a pure shear as in Ref.\cite{19,20,21}, a non-linear elastic transition
can only occur through a major reorganization of the flow structure. In Ref.\cite{19,20,21} the
secondary flow generated above the instability threshold had well expressed turbulent features and
involved irregular fluid motion in a broad range of temporal and spatial scales. That implies an
essentially collective effect of the polymer molecules on the flow, and the polymer solution behaving
as a visco-elastic continuum. If the size of the set-up is reduced to a micro-scale, non-homogeneity
in the polymer concentration and extension of polymer molecules stretched in the flow may become
comparable with the size of the set-up, and larger than size of some of the generated vortices.
Therefore, whether or not the purely elastic instability and elastic turbulence can be reproduced in a
microscopic shear flow is still an open question.

Here we show that a fully developed chaotic flow similar to the elastic turbulence can be generated in
a flat curvilinear microscopic channel with smooth walls and uniform width at arbitrarily low $Re$, if
the working liquid contains a small amount of high molecular weight polymers. The flow is
characterized by significant non-linear growth in resistance, randomly fluctuating velocity field and
chaotic three-dimensional mixing patterns. It is further shown that stirring by the flow results in
efficient mixing in the micro-channel with characteristic mixing length significantly shorter compared
with the "herring-bone" patterning method reported before\cite{14}.  The characteristic mixing times
for solutions of macromolecules are reduced by three to four orders of magnitude compared with
molecular diffusion.

\section{ Materials and Methods.}

\subsection{ Device fabrication.}The micro-channel devices consist of a silicon elastomer
(Sylgard 184 by Dow Corning) chip sealed to a \#1 microscope cover glass. The channel structure of the
chip was fabricated using the technique of soft lithography\cite{1}. First, a negative master mold was
fabricated in UV-curable epoxy (SOTEC micro-systems SU8-1070) by using conventional photolithography.
The epoxy was spun onto a silicon wafer at 1800 rpm for 60 s to create a 100 $\mu$m layer and
patterned by using a high-resolution negative transparency mask. Liquid elastomer was poured on the
mold to a thickness of    $\approx 5$ mm and cured in an 80C oven for 1 hr 30 min. After that the
elastomer was peeled off the mold, trimmed to its final size and liquid feeding ports were punched by
using a 20-gauge luer stub. The patterned side of the chip was bonded to the cover glass by overnight
baking in the 80C oven.

\subsection{Experimental set-up.} A snapshot of our first microfluidic set-up is shown in Fig.1. It has a
uniform thickness $d =100 \mu$m. Its main active element is a curvilinear channel with square
cross-section. It is a chain of 40 identical segments, which are couples of interconnected half-rings
with inner and outer radii $R_i=100 \mu$m and $R_o=200 \mu$m, respectively, Fig.1b. It has the same
proportions as the table-top channel, which was used in the elasticity induced mixing experiments
reported before\cite{21}, but its dimensions are reduced by a factor of 30. Because of the periodic
structure of the channel, it is convenient to use the number of a segment, $N$, starting from the
inlet as a discrete linear coordinate along the channel. The auxiliary rectilinear channel (b) has
width of 90 $\mu$m and total length of about 72.5 mm. Channel (b) and the comparator region (c) serve
to make differential in situ measurements of flux vs. pressure by the method described in
Ref.\cite{22}.

\subsection{Flow control.} The flow in the micro-channels was generated and controlled by pressure
differences between the inlets and the outlet, Fig.1a.  The pressures were generated hydrostatically
using long vertical rails with precise rulers and sliding stages. Working liquids were kept in 30 ml
plastic syringes, which were held upright, open to the atmosphere and connected to the two inlets and
the outlet by plastic tubing with internal diameter of 0.76 mm. The pressure drop in the tubing was
estimated below 1\% of the total. The two syringes feeding the inlets were attached to the sliding
stages. The difference in liquid elevation between these two syringes and the outlet syringe was
measured and adjusted with a precision of about 0.1 mm corresponding to 1 Pa in pressure. Dependence
of the volumetric low rate, Q, in the curvilinear channel on the pressure difference between inlet 1
and outlet, Fig.1, was determined with a relative precision of about 0.5\% using an in-situ
measurement technique described elsewhere\cite{22}. A syringe pump (PHD 2000 by Harvard Apparatus
Inc.) with a 50 ml gastight Hamilton syringe was used for an absolute flow rate calibration.

\subsection{Polymer solutions.} The polymer used was polyacrylamide, PAAm, by Polysciences Inc. with high
molecular weight $M_w=1.8\cdot 10^7$, which was the same polymer sample as in Ref.\cite{21}. It was
dissolved at identical
 concentrations of 80 ppm by weight in two Newtonian solvents with different viscosities. The solvent
  for the low viscosity solution 1 was a 35\% solution of sucrose in water with 1\% of NaCl added to fix
  the ionic contents. The Newtonian viscosity of the solvent, $\eta_s$, was 4.2 mPas at the room temperature
  of 22C. The viscosity of solution 1, $\eta$, was 5.6 mPas at a shear rate of 50 s$^{-1}$, suggesting a dilute
  polymer solution.  The Newtonian solvent for the high viscosity solution 2 was sugar syrup containing
  64.4\% sucrose and 1\% of NaCl in water. The viscosity of solvent 2 was 114 mPas  at 22C, and the
  viscosity of solution 2 was 138 mPas at a shear rate of 2 s$^{-1}$. The polymer relaxation time, $\lambda$,
  of solution 2 measured by phase shift between stress and shear rate in an oscillatory flow regime was 1.1 s.
   The measurements of viscosity and relaxation time were made using a high precision rheometer
   (AR1000 by TA Instruments). Relaxation time of solution 1 was estimated as 0.04 s with the assumption
    that $\lambda$  scales linearly with $\eta_s$\cite{15}.\\
The overlap concentration $c^*$, taken as a concentration at which the viscosity ratio reached
$\eta/\eta_s=2$, was 200 ppm by weight for solvent 2, corresponding to molecular concentration of
$n=8.76\cdot 10^{12}$ cm$^{-3}$. Characteristic size of the polymer coils at rest can be estimated
from this as $n^{-1/3}\approx 0.5\mu$m, and characteristic distance between them at 80 ppm by weight
can be estimated as 0.7 $\mu$m. These estimates are well supported by the data on PAAm taken from the
literature\cite{25}.  One can use the Mark-Houwink scaling relation for PAAm, $[\eta]=6.31\times
10^{-3}M_w^{0.8}$ (in ml/g)\cite{25}, where $[\eta]$ is the intrinsic viscosity of the polymer defined
as $[\eta]=\frac{\eta-\eta_s}{c\eta_s}|_{c\rightarrow 0}$. At molecular weight $M_w=1.8\times 10^7$
one gets $[\eta]=4020$ ml/g. Defining $c^*=1/[\eta]$\cite{15} we obtain $c^*=250$ ppm by weight for an
aqueous solution in a good agreement with the above estimate. Further, we can compare the estimated
size of the PAAm coils with the data obtained from light scattering Ref\cite{25}. Plugging in
$M_w=1.8\times 10^7$ into an interpolation relation from Ref.\cite{15}, one gets $R_g\approx 0.4\mu$m
for the radius of gyration of the coils, which is rather consistent with the above estimate obtained
from $c^*$.

It is worth noting that $c^*=200$ ppm is very close to the value found for  $\lambda$-phage
DNA\cite{26}, which has a comparable molecular weight of $3.1\times 10^7$. Radius of gyration of the
$\lambda$-phage DNA was found to be $0.73\mu$m\cite{27}, which is quite in line with the above
estimate for the PAAm coil size. The full contour length of a PAAm molecule having $M_w=1.8\times
10^7$ and thus consisting of $2.5\times 10^5$ monomers (having molecular weight 71.08 g/mol) can be
estimated as about $50\mu$m, if a monomer length of $0.2$ nm is assumed. It is significantly larger
than the contour length of the $\lambda$-phage DNA (which is equal to about $16\mu$m) and twice
smaller than the micro-channel diameter.

 The experiments on mixing
were carried out with polymer solution 2, and then fluorescent dyes with low diffusivities were added
to the solution and used as tracers. Those were a few different samples of fluorescein-conjugated
Dextran, FITCD, by Sigma with average molecular weights, $M$, varying from 10 kDa to 2 MDa. In spite
of the relatively high molecular weight of FITCD it did not have any measurable influence on the
solution rheology due to high rigidity of the polysaccharide molecules. Diffusion coefficients of the
FITCD samples in water were estimated using the data in Ref.\cite{28}, giving values from $9.1\cdot
10^{-7}$ to $7.4\cdot 10^{-8}$ cm$^2$/s that corresponded to a broad range of biological
macromolecules. The diffusion coefficients in solvent 2 were estimated with assumption that $D\sim
1/\eta_s$, resulting in $D_1=6.6\cdot 10^{-9}$ and $D_2=5.4\cdot 10^{-10}$  cm$^2$/s for 10 kDa and 2
MDa FITCD, respectively.

\subsection{Measuring flow velocity.} Measurements of the flow velocity in the micro-channel were carried
out using custom developed microscopic particle image velocimetry, micro-PIV. The polymer solution was
seeded with 0.2 $\mu$m yellow-green fluorescent beads (Polysciences), and epi-fluorescent imaging of
the flow in the micro-channel, Fig. 1, was made with an inverted microscope (Olympus IMT2) and narrow
band excitation and emission filters in the dichroic filter cube. The objective was a LWD $20\times$,
N.A. = 0.40 , and the images were projected onto a CCD array with $640\times 480$ pixel resolution
(PixelFly camera by PCO, Germany) and digitized to 12 bits. The snapshots of the flow were taken with
even time intervals of 40 ms, and digitally post-processed. Images of out-of-focus particles were
disregarded. Velocity field was found by cross-correlating positions of the particles in two
consecutive snapshots, and the particle velocity vectors were neighbor-validated. (The calculated
velocity field corresponded to the time interval between the two snapshots.) The collected time series
represented velocity values measured at equal distances from interconnections of two half-rings at
$N=35$, and averaged over a $20\times 20\mu$m square region at the middle of the channel and over 4
$\mu$m across the channel mid-plane.

\subsection{Measuring tracer concentration profiles.} Concentrations of the fluorescent dyes, which were
used as passive tracers in the experiments on mixing in the channel, were measured using a commercial
confocal microscope (Fluoview FV500 by Olympus).  It was equipped with a $40\times$ N.A.= 0.85
infinity corrected objective and a 12 bit photomultiplier. The scanning was done at a rate of 56 lines
per second and 512 pixels per line corresponding to a step of 0.18  $\mu$m per pixel .

\section{ Results.}

We measured volumetric flux rate, $Q$, of solution 1 through the curvilinear channel, Fig.1, in a
broad range of applied pressures and calculated the resistance factor, $Z = P /Q$, where $\Delta P$ is
pressure drop per segment. The resistance factor is a constant proportional to viscosity for Newtonian
fluids in linear, low $Re$ regime, and it can be used as a measure of turbulent flow resistance in
large channels at high $Re$.  Fig. 2 shows dependence of $Z$ on $Q$ for solution 1, after $Z$ is
divided by a resistance factor, $Z_0$, found for a Newtonian liquid with the same viscosity, $\eta$.
The ratio $Z/Z_0$ is constant equal to unity in the linear regime at low $Q$.  At $Q$ of about 8.5
nl/s, however, a non-linear transition occurs; $Z/Z_0$ starts to grow and reaches a factor of about
2.8 at high $Q$. The Reynolds number for the channel flow can be defined as $Re=Q\rho/(\eta d)$ . It
was 0.017 at the transition point and 0.14 at highest $Q$, that we tried, so that the inertial effects
were always negligible. The Weissenberg number can be defined as $Wi=4\lambda Q/d^3$, and its
estimated value at the transition point is 1.4, which is comparable with $Wi$ found at purely elastic
transitions in macroscopic set-ups\cite{18,19,21,29}.

In order to get detailed information about structure of the flow above the non-linear transition we
used solution 2 with high viscosity and large  $\lambda$. The polymer relaxation time defines both
characteristic time of changes in the flow and the inverse of flow velocity in the elastic non-linear
regime\cite{15,29}. Therefore, the non-linear flow in solution 2 was expected to be much slower, and
measurements of its characteristics were expected to be more feasible with the standard video
microscopy techniques.  Using the micro-PIV we measured flow velocity in the middle of the curvilinear
channel. Dependence of RMS of fluctuations of the longitudinal component of the flow velocity,
$V_1^{rms}$, (which is the velocity component along the main flow) on $\Delta P$  is shown in Fig.3.
The fluctuations are virtually absent in the linear regime at low pressure. At $\Delta P$ of about 50
Pa, however, $V_1$ starts to fluctuate, and $V_1^{rms}$ begins to grow quickly and non-linearly.  It
can be learned from the inset in Fig.3 that the longitudinal component of the flow velocity,
$\bar{V_1}$, grows linearly at low $\Delta P$, but its growth slows down at the same critical $\Delta
P$ of about 50 Pa, which is another evidence for a non-linear elastic flow transition taking place. At
the transition point, the average longitudinal flow velocity, estimated from the micro-PIV
measurements in different points across the channel, was $\bar{V_1}\approx 80 \mu$m/s. That gives
estimates of $Q\approx 0.8$ nl/s, $Re\approx 8\cdot 10^{-5}$ and $Wi\approx 3.5$ for the elastic
non-linear transition. This value of the Weissenberg number is very close to $Wi_c\approx 3.2$ found
for the transition to chaotic flow in the table-top curvilinear channel in Ref.\cite{21}

A typical time series of $\bar{V_1}$  above the transition is shown in Fig.4a. The velocity is
strongly fluctuating and its time dependence has well expressed chaotic appearance. The chaotic
character of the velocity fluctuations is confirmed by analysis of its time correlations. The velocity
autocorrelation function shown in Fig. 4b does not have distinct peaks and decays uniformly.

For the experiments on mixing the design of the micro-channel was slightly modified to enable side by
side injection of two streams of solution 2, one with and one without FITCD, to the channel inlet,
Fig.5a. Apart from  $c_0=280$ ppm by weight of FITCD added to one of them the polymer solutions were
identical and were injected at equal flow rates by careful adjustment of the driving pressures, Fig.
5a. The set-up was first tested with the plain solvents without PAAm added. The flow appeared laminar
at all $\Delta P$ that we applied, and the interface between the streams with and without FITCD
remained smooth and sharp along the whole channel with only minor smearing by diffusion, Fig. 5b.

The situation was similar with the polymer solutions in the linear regime at low $\Delta P$. However,
when the driving pressure was raised above the non-linear transition threshold, fluctuating flow
velocity produced significant stirring and complex and chaotically changing tracer concentration
profiles, Fig. 5c, d. We studied mixing in the channel in detail at $\Delta P=134$ Pa, corresponding
to a flow rate about twice above the non-linear transition threshold (cf. Ref.\cite{21}) and
$\bar{V_1}$ of about 173 $\mu$m/s. Variation of the tracer concentration profiles with time at
different distances from the inlet is illustrated by the space-time plots in Fig. 6a,b. One can
observe that the tracer concentration appears to fluctuate quite randomly without any apparent scale
in time or space. Next, one can see in Fig. 6a, taken at $N=12$, that the left side of the channel,
where the tracer was initially injected, looks much brighter and has much higher average concentration
of the tracer. Although also noticeable in Fig.6b taken further downstream, at $N=18$, this feature is
clearly weaker there. Thus, stirring by the fluctuating velocity field seems to create a more
symmetric distribution of the tracer between the two sides of the channel. In order to validate this
observation, we measured time averages of the tracer concentration, $\bar{c}$,  at different positions
across the channel and at different $N$, Fig.7a. One can see that the cross-channel distribution of
$\bar{c}/c_0$ close to the inlet, at $N=7$, is strongly influenced by the asymmetric conditions at the
channel entrance. As one can learn from the curve at $N=11$, however, the imprint of the initial
conditions is clearly fading as the liquid advances downstream and being stirred. Further downstream,
at $N=41$, asymmetry in the tracer distribution introduced by the initial conditions disappears
completely. Fading of the initial condition influence with time and restoration of symmetry in flow in
statistical sense are both distinct features of chaotic and turbulent flows. Therefore, the curves in
Fig. 7a provide further evidence for truly chaotic nature of the flow in the micro-channel.

    A natural parameter characterizing inhomogeneity of the mixture is standard deviation of the
instantaneous local tracer concentration, $c$, from its overall average value $\langle c\rangle=c_0$.
It is convenient
 to divide it by $\langle c\rangle$  and to introduce a dimensionless standard deviation $c_{std}=\sqrt{\langle
 (c-\langle c\rangle)^2\rangle}/\langle c\rangle$. At the channel entrance $c_{std}$
is equal to unity, and it becomes zero, when the liquid is perfectly mixed and homogeneous.
 Dependence of $c_{std}$  on $N$ for two tracers with diffusion coefficients of $D_1=6.6\cdot 10^{-9}$ and
  $D_2=5.4\cdot 10^{-10}$
cm2/s is shown in Fig.7b in semi-logarithmic scale. One can see that the mixture becomes increasingly
homogeneous as the liquid advances downstream, and the parameter $c_{std}$  decays exponentially with
$N$ for the both tracers. An exponential decay was also found in the case of elastic turbulence in a
macro-channel \cite{21}, and it agrees very well with theoretical predictions for the so-called
Batchelor regime\cite{30,31,32,33} of mixing. The latter corresponds to a flow, which is chaotic in
time but essentially "smooth" in space, in the sense that small eddies are rare, and the main
contribution to mixing comes from the largest eddies having the size of the whole system.

The rates of the exponential decay for the two dependencies shown in Fig. 7b were 0.217 and 0.137,
corresponding to $\Delta N$ of 4.61 and 7.30, and lengths  $L_1=4.34$ mm and  $L_2=6.88$ mm,
respectively. The latter can be considered as characteristic mixing distances along the channel for
the two tracers. Characteristic mixing times, which can be estimated as $t_{mix}=L/\bar{V}$, are then
found to be $t_{mix,1}=25$ s and   $t_{mix,2}=40$ s, respectively. In the absence of active stirring,
mixing would only occur through molecular diffusion, with characteristic diffusion times across the
channel given by $t_{diff}=d^2/D$ resulting in  $t_{diff,1}=1.5\cdot 10^4$ s and $t_{diff,2}=1.9\cdot
10^5$ s, respectively. Therefore, the stirring produced by the chaotic flow in the channel reduces the
mixing times for the FITCD macromolecules by three to four orders of magnitude.

\section{ Discussion.}

We studied flow of two dilute polymer solutions in a microscopic curvilinear channel. We observed a
non-linear transition to occur in the flow at $Re\approx 1.7\cdot 10^{-2}$, $Wi\approx 1.4$, and
$Re\approx 8\cdot 10^{-5}$, $Wi\approx 3.5$, for solution 1 and solution 2, respectively. The very
small and different values of $Re$, and comparable, order of unity values of $Wi$ at the transition
threshold for the two solutions indicate that the transition is of purely elastic nature and can occur
at arbitrarily low $Re$. A possible explanation of the difference in estimated critical $Wi$ for the
two polymer solutions is that the actual polymer relaxation time of solution 1 may be significantly
higher than the estimate based on the assumption that $\lambda\sim\eta_s$.\\
 A major global feature of
the flow above the non-linear elastic transition is fast growth of the resistance, Fig.2. It increases
by up to a factor of 2.8 above the resistance for a Newtonian fluid with the same shear flow
viscosity, $\eta$, at the same $Q$. Since $Re$ is very low, the whole growth of the flow resistance
should be due to increase in the elastic polymer stresses\cite{20}. In a pure shear flow with
$\dot{\gamma}=240$ s$^{-1}$, corresponding to $Q=60$ nl/s (and $Wi\approx 10$), viscosity ratio for
polymer solution 1 was $\eta/\eta_s=1.22$. Hence, the average increase in the polymer shear stresses
due to the secondary flow at $Q=60$ nl/s (Fig.2) can be estimated as a factor of about 11. Suggesting
linear deformation of the flexible polymer molecules in a shear flow at low to moderate $Wi$\cite{34},
we can estimate extension of the molecules at $Wi=10$ as $10\cdot 0.5  \mu$m = $5 \mu$m. (Flexible
polymer molecules were found to extend linearly with the shear rate, until the extension reached
15-20\% of their full length\cite{34}.) Additional extension of the polymers due to the irregular
secondary flow, as suggested by the stress growth by the factor of $11$, can be estimated as a factor
of $\sqrt{11}$ (suggesting that the secondary flow results in an isotropic polymer unravelling and the
stress grows as a square of the polymer extension\cite{15}). That brings characteristic size of the
extended polymer molecules to $20\mu$m range, less than an order of magnitude smaller than the
diameter of the channel, $d=100\mu$m.

The growth of the elastic stresses due to the fluctuating secondary flow can only occur as a result
of significant reorganization of the flow structure and spontaneous generation of regions with strong
 extensional flow, as in the case of the elastic turbulence\cite{19,20}. (In shear flows polymer contribution
  to resistance increases slower than linearly with the shear rate, which is called shear thinning.)
  This suggestion is quite corroborated by the direct flow velocity measurements in solution 2 above the
   non-linear transition. Velocity is found to be strongly fluctuating, Figs.3,4, with RMS of the
   fluctuations reaching as much as 10\% of the mean longitudinal velocity at the center of the channel,
    Fig.3, just as in the table-top channel in Ref.\cite{21}. The velocity appears to vary randomly in time,
    Fig.4a, and its auto-correlation function decays quite quickly and does not have any distinct peaks,
     Fig.4b. These all are clear indications of chaotic nature of the flow in the channel in the
     non-linear regime at high $Wi$. Another evidence of chaos comes from the experiments on mixing,
     Figs.6,7. Asymmetry in distribution of $\bar{c}/c_0$  across the channel, which is imposed by the
     conditions
     at the channel entrance, decays with the distance from the entrance, Fig7a. Further, $c_{std}$
     exponentially
     decays with $N$, as it is supposed to be in the chaotic Batchelor regime of mixing\cite{21,32,33}.

Experimental results presented in Figs.2-4 and Figs.6,7 and discussed above are fully consistent with
a suggestion that the regime of elastic turbulence\cite{19} is being realized in the micro-channel. In
fact, the results in Figs.2-4 and Figs.6,7 agree rather well with the measurements in macroscopic
systems reported before \cite{19,20}. We find it rather remarkable that although the size of extended
molecules may become comparable with the channel diameter, it does not seem to cause any significant
new effects in the polymer solution dynamics. An essential feature of turbulent flows is a broad range
of spatial scales, at which fluid motion is excited. Unfortunately, limited resolution of the
micro-PIV technique did not allow us to explore properties of the flow down to sufficiently small
spatial scales and to obtain conclusive data about its spatial structure. Therefore we can only refer
to the flow in the micro-channel as being chaotic.

In the experiments with passive tracers, we found that stirring by the chaotic flow results in
efficient mixing in the channel, Figs.5-7. So, mixing times for the macromolecules, which we used as
tracers, were reduced by three to four orders of magnitude compared with passive molecular diffusion,
Fig.7b. It is worth noting that the characteristic distance of $\Delta N=7.3$ required for mixing of
low diffusivity FITCD is comparable with $\Delta N\approx 15$ found in the chaotic flow in the
table-top channel Ref.\cite{21}. We further notice that whereas the diffusion time scaled as $1/D$,
the time of mixing in the chaotic flow depended on $D$  very
 weakly. It increased by only 60\% between the lower and upper curve in Fig.7b (from   $t_{mix,1}=25$ s
 and   $t_{mix,2}=40$ s),
  whereas  $D$ dropped by a factor of 12 (from $D_1=6.6\cdot 10^{-9}$ to $D_2=5.4\cdot 10^{-10}$  cm2/s).
  The weak dependence
  of  $t_{mix}$ on $D$  is quite consistent with theoretical predictions for chaotic flows and the
  Batchelor regime
  \cite{30,31,32,33}. It implies that the elasticity induced chaotic stirring in the micro-channel can be
  efficiently used for mixing of liquids with additives of any low diffusivity, such as large molecules of
  DNA, viruses, particles, and possibly living cells.

The relatively long times of mixing in the chaotic flow of solution 2 are due to high viscosity of the
solvent, large $\lambda$  and low flow velocity at the elastic instability threshold. In a much more
practically relevant case of water based solutions with the viscosity on the order of 0.001 Pas,
$\lambda$ is expected to be about hundred times lower and the flow velocities in the chaotic regime
should be about hundred times higher\cite{15,29}. Suggesting qualitatively similar flow
conditions\cite{29}, we expect characteristic mixing times for the water based solutions to be on 100
ms scale. Since diffusion coefficients for macromolecules are proportional to $1/\eta_s$\cite{15}, the
ratio between $t_{diff}$  and $t_{mix}$ for the chaotic flow in the water based solutions should be in
the same range of $10^3$ to $10^4$.

The characteristic mixing length along the channel for water based solutions should also remain in the
same range of about 4-7 mm. It is about 2-3 times shorter than characteristic lengths in the
"herring-bone" patterned channel of Ref.\cite{14} at comparable values of Peclet number, $Pe=Vd/D$.
Although the method of mixing by elasticity induced chaotic flow necessarily requires addition of
polymers to the working liquid, it may be quite practical for many biochemical assays, taking into
account the very low concentration of the high molecular weight polymers used. It does not rely on any
special patterning of the micro-channels and should be readily compatible with rounded channel
profiles used for integrated micro-valves and peristaltic pumps\cite{35}, allowing efficient mixing in
closed loop microscopic flows\cite{36}.\\

We are grateful to M. Chertkov and V. Lebedev for many useful and illuminating discussions and to D.
Mahalu for valuable help with the microfabrication. One of us (T. B.) thanks V. Kiss for his support
in the confocal set-up measurements. This work is partially supported by an Israel Science Foundation
grant, Binational US-Israel Foundation, and by the Minerva Center for Nonlinear Physics of Complex
Systems.

\begin{figure}

\caption{(A)  Photograph of the micro-fluidic device. The micro-channels were filled with ink for
better contrast. (B) Magnified image of a section of the functional curvilinear channel.}

\label{figa}
\end{figure}

\begin{figure}

\caption{Dependence of normalized resistance, $Z/Z_0$, in flow of solution 1 through the curvilinear
channel on the volumetric flow rate, $Q$, in semi-logarithmic scale.}

\label{figb}
\end{figure}

\begin{figure}

\caption{Dependence of RMS of fluctuations of the longitudinal component of flow velocity, $V_1^{rms}$
, on pressure drop per segment, $\Delta P$, in the center of the micro-channel  for solution 2. Inset:
Dependence of time average of the longitudinal flow velocity, $\bar{V_1}$ , in the center of the
micro-channel on $\Delta P$.}

\label{figc}
\end{figure}

\begin{figure}

\caption{(A) Time series of the longitudinal flow velocity, $V_1$ , in the center of the micro-channel
at $\Delta P=100$Pa. (B) Auto-correlation function for $V_1$ based on about 6000 individual velocity
measurements. }

\label{figd}
\end{figure}

\begin{figure}

\caption{ (A) Epi-fluorescent micro-photograph of the entrance area of a micro-channel used in
experiments on mixing. Wide triangular region in front of a curvilinear channel allows to adjust equal
flow rates for polymer solutions with (from below) and without FITCD. (B) Confocal photograph of flow
in the micro-channel without polymers added. Right wall of the channel is shown by a dotted line from
below. (C) Confocal image of mixing in chaotic flow in the micro-channel with solution 2 in it. (D)
Confocal image of cross-section of the micro-channel with chaotic flow in solution 2.}

\label{fige}
\end{figure}

\begin{figure}

\caption{ Space-time plots of FITCD distribution across the channel taken at (A) $N=12$ and (B)
$N=18$. Confocal scanning was done along the same line across the channel in the mid-plane at equal
distances from half-ring interconnections, with even time intervals of 0.0177 sec. Profiles of FITCD
concentration in consecutive moments of time are plotted from top to bottom.}

\label{figf}
\end{figure}

\begin{figure}

\caption{(A) Time average of FITCD concentration, $\bar{c}$ , divided by $c_0$, as a function of
position, $x$, along a line across the channel taken at different distances from the inlet: $N=7$,
$N=11$, and  $N=41$. The lines across the channel were in the mid-plane at equal distances from
half-ring interconnections, just as for the space-time plots in Fig.6. (B) Standard deviation of FITCD
concentration from its average value, $c_{std}$, as a function of distance, $N$, from the inlet for
two kinds of FITCD with average molecular weights of 10 kDa (triangles) and 2 MDa (squares).}

\label{figg}
\end{figure}

\end{multicols}{2}


\begin{references}

\bibitem{1} { Y. N. Xia, G.M. Whitesides, {\sl Ann. Rev. Material Sc.} {\bf28}, 153(1998).}
\bibitem{2} {G. M. Whitesides, A.D. Stroock, {\sl Physics Today} {\bf 54}(6), 42 (2001). }
\bibitem{3} { C. L. Hansen, E. Skordalakes, J.M. Berger, \&  S.R. Quake, {\sl PNAS} {\bf 99}, 16531 (2002).}
\bibitem{4}{ N. L. Jeon,  H. Baskaran, S.K.W. Dertinger,  G.M. Whitesides, L. Van de Wate, \& M. Toner,
{\sl Nature Biotech.} {\bf 20}, 826 ( 2002).}
\bibitem{5}{S. Takayama, E. Ostuni, P. LeDuc, K. Naruse, D.E. Ingber, \& G.M. Whitesides,
{\sl Nature} {\bf 411}, 1016 (2001).}
\bibitem{6}{ H. B. Mao,  P.S. Cremer, \& M.D. Manson,  {\sl PNAS} {\bf 100}, 5449 (2003).}
\bibitem{7}{ H. P. Chou, C. Spencer, A. Scherer, \& S. Quake,  {\sl PNAS} {\bf 96}, 11 (1999).}
\bibitem{8}{  L. D. Landau, \&  E.M. Lifshitz, {\sl Fluid Mechanics} (Pergamon Press)(1987).}
\bibitem{9}{  L. Reyes,  J. Bert,  J. Fornazero, R. Cohen, \&  L. Heinrich, {\sl Colloids and
surfaces} {\bf B 25}, 99 (2002). }
\bibitem{10}{  M. H. Oddy,  J. G. Santiago, \& J. C. Mikkelsen,  {\sl Anal. Chem.} {\bf 73}, 5822 (2001).}
\bibitem{11}{J. H. Tsai,  \&  L.W. Lin, {\sl Sens. Actuat.} {\bf A 97}, 665 (2002).}
\bibitem{12}{  D. Therriault, S.R. White, \& J.A. Lewis,  {\sl Nature Materials} {\bf 2}, 265 (2003).}
\bibitem{13}{  R. A. Vijayendran, K.M. Motsegood,  D.J. Beebe, \& D.E.Leckband, {\sl Langmuir } {\bf 19},
1824 (2003).}
\bibitem{14}{ A. D. Stroock, S. K. Dertinger, A. Ajdari, I. Mezic, H. A. Stone, \&
G. M. Whitesides,  {\sl Science} {\bf 295}, 647 (2002).}
\bibitem{15}{ R. B. Bird, Ch. Curtiss,  R.C. Armstrong, \& O. Hassager  {\sl Dynamics of Polymeric
Liquids}, (Wiley, New York)(1987).}
\bibitem{16}{ V. Tirtaatmadja \& T. Sridhar, {\sl J. Rheology} {\bf 37}, 1081 (1993).}
\bibitem{17}{ D. V. Boger, K. Walter, {\sl Rheological phenomena in focus} (Elsevier) (1993). }
\bibitem{18}{ R. G. Larson,  E.S.G. Shaqfeh, \& S.J. Muller,  {\sl J. Fluid Mech.} {\bf 218}, 573 (1990).}
\bibitem{19}{ A. Groisman, \&  V. Steinberg, {\sl Nature} {\bf 405}, 53 (2000).}
\bibitem{20}{ A. Groisman, \&  V. Steinberg, {\sl Phys. Rev. Lett.} {\bf 86}, 934 (2001).}
\bibitem{21}{ A. Groisman, \&  V. Steinberg, {\sl Nature} {\bf 410}, 905 (2001).}
\bibitem{22}{ A. Groisman,  M. Enzelberger, \& S.R. Quake,  {\sl Science} {\bf 300}, 955 (2003).}
\bibitem{23}{P. G. DeGennes, {\sl J. Chem. Phys.} {\bf 60}, 5030 (1974). }
\bibitem{24}{  T. T. Perkins, D.E. Smith, \& S. Chu,  {\sl Science} {\bf 276}, 2016 (1997).}
\bibitem{25}{W.-M. Kulicke, M. Kotter, \& H. Grager, in {\sl Advances in Polymer Science 89}, {\sl
Polymer characterization/Polymer solutions}, (Springer-Verlag, Berlin), 1989.}
\bibitem{26}{ P.J. Shrewbury, S. J. Muller, \& D. Liepmann, {\sl Biomedical Microdevices} {\bf 3}, 225 (2001).}
\bibitem{27}{ D.E. Smith, T. Perkins, \& S. Chu,  {\sl Macromolecules} {\bf 29}, 1372 (1996).}
\bibitem{28}{ C. Wu, {\sl Macromolecules} {\bf 26}, 3821 (1993).}
\bibitem{29}{ A. Groisman, \& V. Steinberg, {\sl Europhys. Lett.} {\bf 43}, 165 (1998).}
\bibitem{30}{G. K. Batchelor,  {\sl J. Fluid Mech.} {\bf 5}, 113 (1959).}
\bibitem{31}{ M. Chertkov, G. Falkovich, I. Kolokolov, \& V. Lebedev, {\sl Phys. Rev. E} {\bf 51}, 5609 1995).}
\bibitem{32}{T. D. Son, {\sl Phys. Rev E} {\bf 59}, R3811 (1999).}
\bibitem{33}{ E. Balkovsky, \& A. Fouxon, {\sl Phys. Rev E} {\bf 60}, 4164 (1999).}
\bibitem{34}{ D.E. Smith, H. P. Babcock, \& S. Chu, {\sl Science} {\bf 283}, 1724 (1999).}
\bibitem{35}{ M. A. Unger, H.P. Chou, T. Thorsen,  A. Scherer, \& S.R. Quake, {\sl Science}
{\bf 288}, 113 (2000).}
\bibitem{36}{H. P. Chou, M.A. Unger, \& S.R. Quake, {\sl Biomedical Microdevices} {\bf 3}, 323 (2001).}
\end{references}
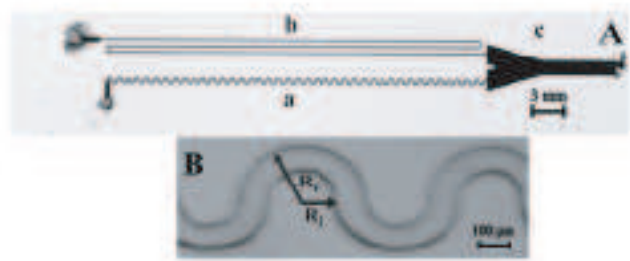
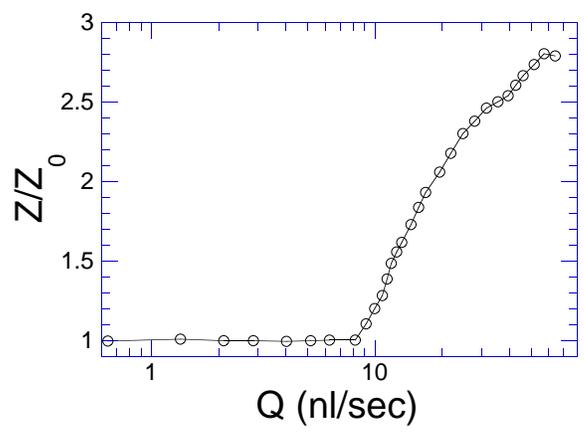
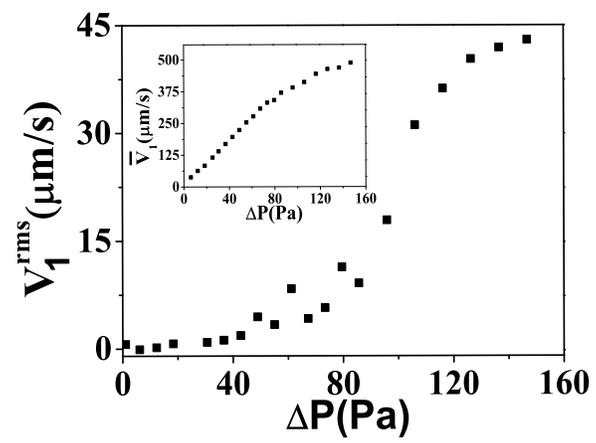
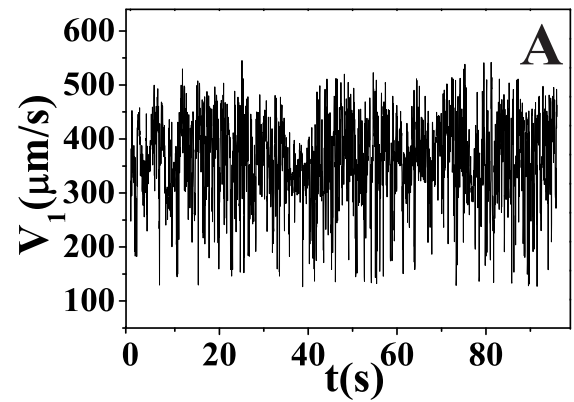
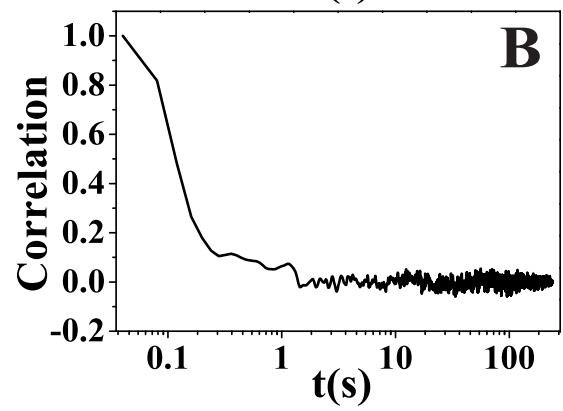
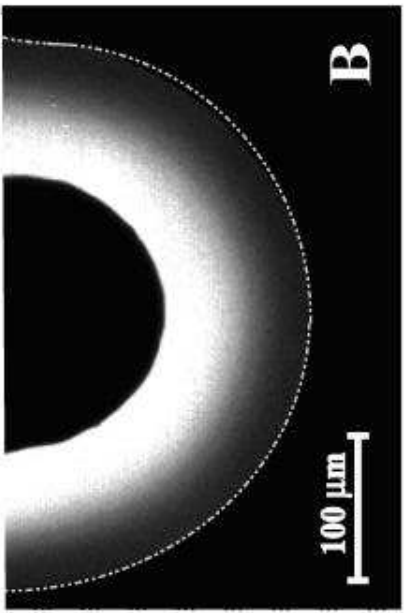
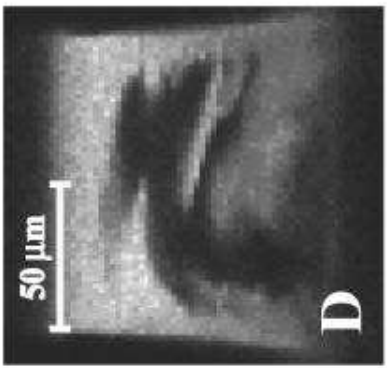
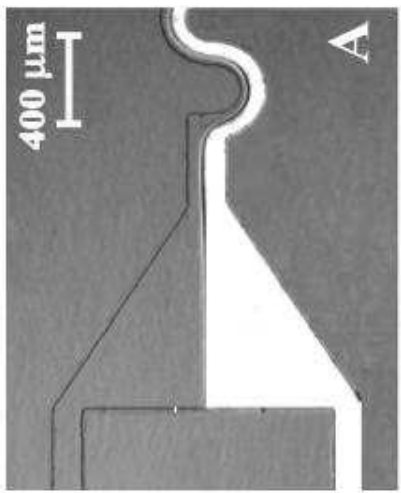
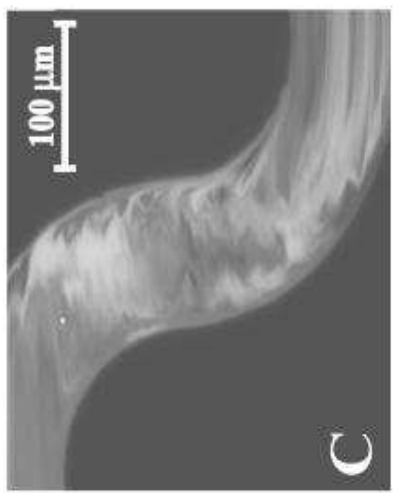
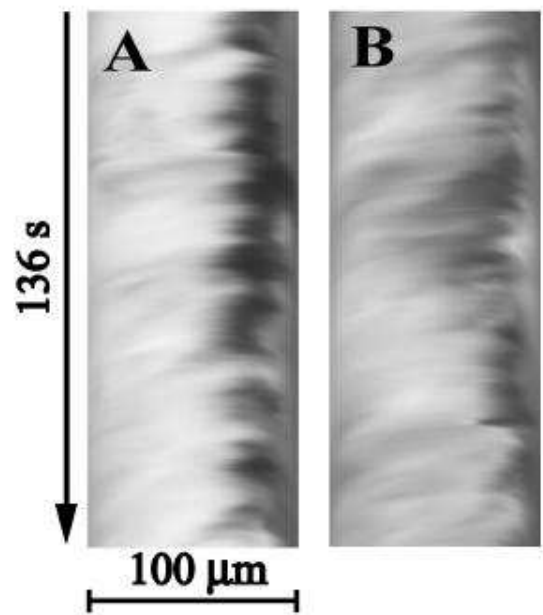
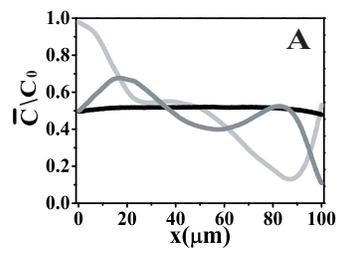
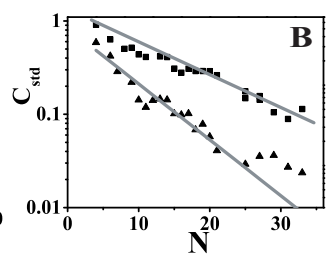
\end{document}